\documentclass[12pt,preprint]{aastex}
\usepackage{graphicx}
\usepackage{epstopdf}
\usepackage{lineno}
\renewcommand\linenumberfont{\normalfont\sffamily\small}
\usepackage{xcolor}

\newcommand{\kms}{km s$^{-1}$}
\newcommand{\solar}{\ifmmode_{\sun}\;\else$_{\sun}$\fi}
\newcommand{\HI}{H$\,${\sc i}}
\newcommand{\sigdens}{M\solar\ pc$^{-2}$}

\newcommand{\optsb}{mag arcsec$^{-2}$}
\newcommand{\dirr}{dIrr}

\begin{document}

\title{Ultra-deep imaging of nearby dwarf irregular galaxies: stellar haloes and disk structure}

\author{Deidre A.\ Hunter}
\affil{Lowell Observatory, 1400 West Mars Hill Road, Flagstaff, Arizona 86001 USA; dah@lowell.edu} 
\and
\author{Bruce G.\ Elmegreen}
\affil{Katonah, NY, USA}

\begin{abstract}
We have examined the stellar structure of 10 nearby, low stellar mass ($10^6$ to $6 \times10^7$ M\solar)
dwarf irregular galaxies by fitting ellipses as a function of surface brightness
on ultra-deep $V$ images. These are compared to far ultraviolet images as tracers of the star formation.
We find that the often asymmetrical distribution of large patches of star formation activity in dwarfs, 
even out to low disk surface brightness levels,
skews the broad-band optical isophotes in these galaxies.
We also looked for evidence of the presence of a stellar halo. Possible hints of such are found in 
several galaxies from irregularities in the ellipses, but a stack of seven of the galaxies shows a
pure exponential out to a $V$ surface brightness of 32.3 \optsb\ where the stellar surface density is $0.0013\pm0.0011$ \sigdens.
The extended stellar component, most likely a disk structure, is probably due to internal evolutionary 
processes rather than external accretion.
The $UBVI$ colors of the annuli are consistent with ages of 1-6 Gyr for the far outer stellar disk.
\end{abstract}


\section{Introduction} \label{sec-intro}

Ultra-deep imaging in $V$-band of dwarf irregular (\dirr) galaxies has shown that 
dwarfs can have extended stellar disks, most often exponential, that extend to many disk scale lengths. 
\citet{pilot} traced the exponential disks of 5 \dirr\ galaxies to 3-8.5 disk scale lengths, corresponding
to a $V$-band surface brightness $\mu_V$ of 29.5 \optsb,
and \citet{hunter25} examined the far outer stellar disk to 29 \optsb\ in 10 additional \dirr\ galaxies.
Using stellar population analysis, \citet{Saha10} detected the stellar disk of the LMC to 12 disk scale lengths,
corresponding to an $I$-band surface brightness of 34 \optsb.
\citet{sextansab} also used star counts to trace the stellar disks of \dirr\ galaxies Sextans A and Sextans B to 
a $\mu_V$ of 31 \optsb, and \citet{hidalgo03} traced stars in DDO 165 and DDO 181 to a $\mu_R$ of 28-29 \optsb.
\citet{Tau24} found RR Lyrae stars up to four times the half-light radius in 10 ultrafaint dwarfs.
\citet{d216} used star counts in DDO 216 to detect stars down to a surface density of 0.0005 stars pc$^{-2}$,
\citet{Kado} detected an extended stellar component, mostly exponential,
in a large sample of more massive dwarfs ($\ge 10^{8.5}$ M\solar), and \citet{Jang} have studied the
extended stellar component of NGC 300.

In addition to extended stellar disks, stellar haloes are predicted by simulations and are seen by some observations.
Extended stellar distributions have been interpreted as 
haloes observed around, for example, DDO 69 \citep{LeoA18}, NGC 3109 \citep{n3109},
WLM \citep{WLM}, IC 1613 \citep{ic1613}, and possibly DDO 187 \citep{d187}. 
(See \citet{Stinson} for additional references).
Although stellar haloes can be built from the disruptions of a galaxy-galaxy interaction, as seen
in the streams and shells around NGC 300 \citep[][see also \citet{Spitzer,Bullock,Deason}]{n300}, 
here we are exploring haloes that might be more
common, having been built from normal evolutionary processes  
\citep[as inferred from  Figure 1 of][]{Stinson}.
In simulations, for example, 
\citet{elbadry16} \citep[see also,][]{benavides,governato} found that \dirr\ galaxies often undergo ``breathing modes,'' puffing up
after star formation to a radius that is up to twice the equivalent radius and then shrinking. 
This dynamical evolution is expected to scatter stars into the far outer disk. 
This might also populate a stellar halo around the galaxies. 
Simulations by \citet{Stinson} explore the effects of star formation activity contracting in radius with time 
\citep[as observed by][]{zhang12}, 
scattering of stars, and stars ejected by supernova-driven shocks. These processes would be particularly
important in low mass \dirr\ galaxies. Their simulations produced extended multi-component (broken) exponentials
with an old stellar population in the far outer regions.

Here, we use the ultra-deep imaging of \citet{hunter25} to look at the structure of the stellar disks
as a function of surface brightness in $V$, from the central regions to the far outer regions. 
Furthermore, we examine these data for evidence of a transition from the stellar disk to a stellar halo. 
The galaxies discussed here are relatively isolated and not obviously interacting, so that we
are most likely exploring inherent evolutionary processes. The exceptions to this are discussed below.
In Section \ref{sec-obs} we describe the galaxy sample and imaging data. 
In Section \ref{sec-analysis} we show the fitting of ellipses to surface brightness levels and the
position angles (PAs) and minor-to-major axis ratios ($b/a$) as a function of surface brightness.
In Section \ref{sec-discuss} we discuss the effects of star formation on the structure of the inner stellar disk
relative to the outer disk, evidence for external influences including mergers on three galaxies, and the search for
stellar haloes in these systems.
A summary of our findings is given in Section \ref{sec-summary}.

\section{Data} \label{sec-obs}

\subsection{Galaxy sample and data}

The 10 \dirr\ galaxies observed in the ultra-deep imaging survey of \citet{hunter25} are from the 
Local Irregulars That Trace Luminosity Extremes, The \HI\ Nearby Galaxy Survey (LITTLE THINGS) sample.
The ultra-deep galaxies and pertinent properties are listed in Table \ref{tabgalaxies}; additional properties can be found in \citet{lt}.
One of the 10, NGC 3738, is a Blue Compact Dwarf (BCD), and as such, has a  higher rate of star formation and more centrally concentrated
structure than typical \dirr\ galaxies.
LITTLE THINGS\footnote[1]{The original survey was funded in part by the National Science Foundation through grants AST-0707563, AST-0707426, AST-0707468, and AST-0707835 to US-based LITTLE THINGS team members and supported with generous technical and logistical support from the National Radio Astronomy Observatory.}
is a multi-wavelength survey put together for the purpose of
determining what drives star formation in \dirr\ galaxies \citep{lt}.
The LITTLE THINGS dwarfs are relatively nearby ($\leq$10.3 Mpc),
contain gas so they have the potential for star formation, and are not companions to larger galaxies. 
The sample was also chosen to cover a large range in dwarf galactic properties such as
star formation rate (SFR) and absolute magnitude.
The ultra-deep sample galaxies are at 0.7 Mpc to 8.7 Mpc distances
and sample the range of other parameters of the LITTLE THINGS galaxies.

The LITTLE THINGS ancillary data include far-ultraviolet (FUV) images
obtained with the NASA {\it Galaxy Evolution Explorer} satellite 
\citep[{\it GALEX\footnote[2]{{\it GALEX} was operated
for NASA by the California Institute of Technology under NASA contract NAS5-98034.}};][]{galex}
to trace star formation over the past 100 Myr.
We use the data from the data release described by \citet{gr2-3}.

\begin{deluxetable}{lccccrc}
\scriptsize
\tablecaption{Galaxy Sample. \label{tabgalaxies}}
\tablewidth{0pt}
\tablehead{
\colhead{} & \colhead{R.A.\ (2000.0)\tablenotemark{a}} & \colhead{Decl.\ (2000.0)\tablenotemark{a}} & \colhead{$D$\tablenotemark{b}} & \colhead{$M_V$\tablenotemark{c}} & \colhead{$R_D$\tablenotemark{d}} & \colhead{} \\
\colhead{Name} & \colhead{(hh:mm:ss.s)} & \colhead{(dd:mm:ss)} & \colhead{(Mpc)} & \colhead{(mag)} & \colhead{(kpc)} & \colhead{$E(B-V)_f$\tablenotemark{e}} \\
}
\startdata
DDO 43   &  7:28:17.8  &  40:46:13  &   7.8  &    -15.1  & $0.87\pm0.10$ & 0.05 \\
DDO 46   &  7:41:26.6  &  40:06:39  &   6.1  &    -14.7  & $1.13\pm0.05$ & 0.05 \\
DDO 47   &  7:41:54.8  &  16:48:16  &   5.2  &    -15.5  & $1.34\pm0.05$ & 0.02 \\
DDO 69   &  9:59:25.0  &  30:44:42  &   0.8  &    -11.7  & $0.19\pm0.01$ & 0.00 \\
DDO 187  & 14:15:56.7  &  23:03:19  &   2.2  &    -12.7  & $0.37\pm0.06$ & 0.00 \\
DDO 210  & 20:46:52.0  & -12:50:51  &   0.9  &    -10.9  & $0.16\pm0.01$ & 0.03 \\
DDO 216  & 23:28:35.0  &  14:44:30  &   1.1  &    -13.7  & $0.52\pm0.01$ & 0.02 \\
F564-V3  &  9:02:53.9  &  20:04:29  &   8.7  &    -14.0  & $0.63\pm0.09$ & 0.02 \\
LGS3     &  1:03:55.2  &  21:52:39  &   0.7  &     -9.7  & $0.14\pm0.01$ & 0.04 \\
NGC 3738 & 11:35:49.0  &  54:31:23  &   4.9  &    -17.1  & $0.77\pm0.01$ & 0.00 \\
\enddata
\tablenotetext{a}{Galaxy center positions from \citet{lt}.}
\tablenotetext{b}{Distances ($D$) are from \citet{D09}, \citet{D02}, \citet{K02}, \citet{K03}, \citet{K04}, \citet{K06}, \citet{meschin09}, and \citet{miller01}.}
\tablenotetext{c}{Absolute $V$ magnitudes ($M_V$) are from \citet{lt}.}
\tablenotetext{d}{Stellar disk scalelengths ($R_D$) are from \citet{kim13}.}
\tablenotetext{e}{Foreground reddening from the Milky Way $E(B-V)_f$ are from \citet{bh84}.}
\end{deluxetable}

The ultra-deep imaging was done with Lowell Observatory's 4.3 m Lowell Discovery Telescope (LDT)
and the Large Monolithic Imager (LMI\footnote[3]{ Funded by NSF with grant AST-1005313 to PI Philip Massey.}).
The camera was specifically designed for low surface brightness projects.
The field of view of the CCD is 12.5\arcmin\  $\times$ 12.5\arcmin.
Each galaxy was observed for a total of 10/5/2.5/2.5 hours
in $U/B/V/I$ filters, aiming to reach individually a $\mu_V$ of 30 mag/arcsec$^2$
with smoothing over approximately 10 pixels (scale of 0.24\arcsec). 
The data were taken only on moonless nights, and sky in typical $V$ images had a surface brightness
of 21.3$\pm$0.05 \optsb.
The $I$-band images were de-fringed, dark-sky flats were applied to $U$-band images,
the images in each filter for each galaxy were co-added, and photometric calibrations were determined.
For more details see \citet{hunter25}.

\begin{figure}[t!]
\epsscale{0.65}
\plotone{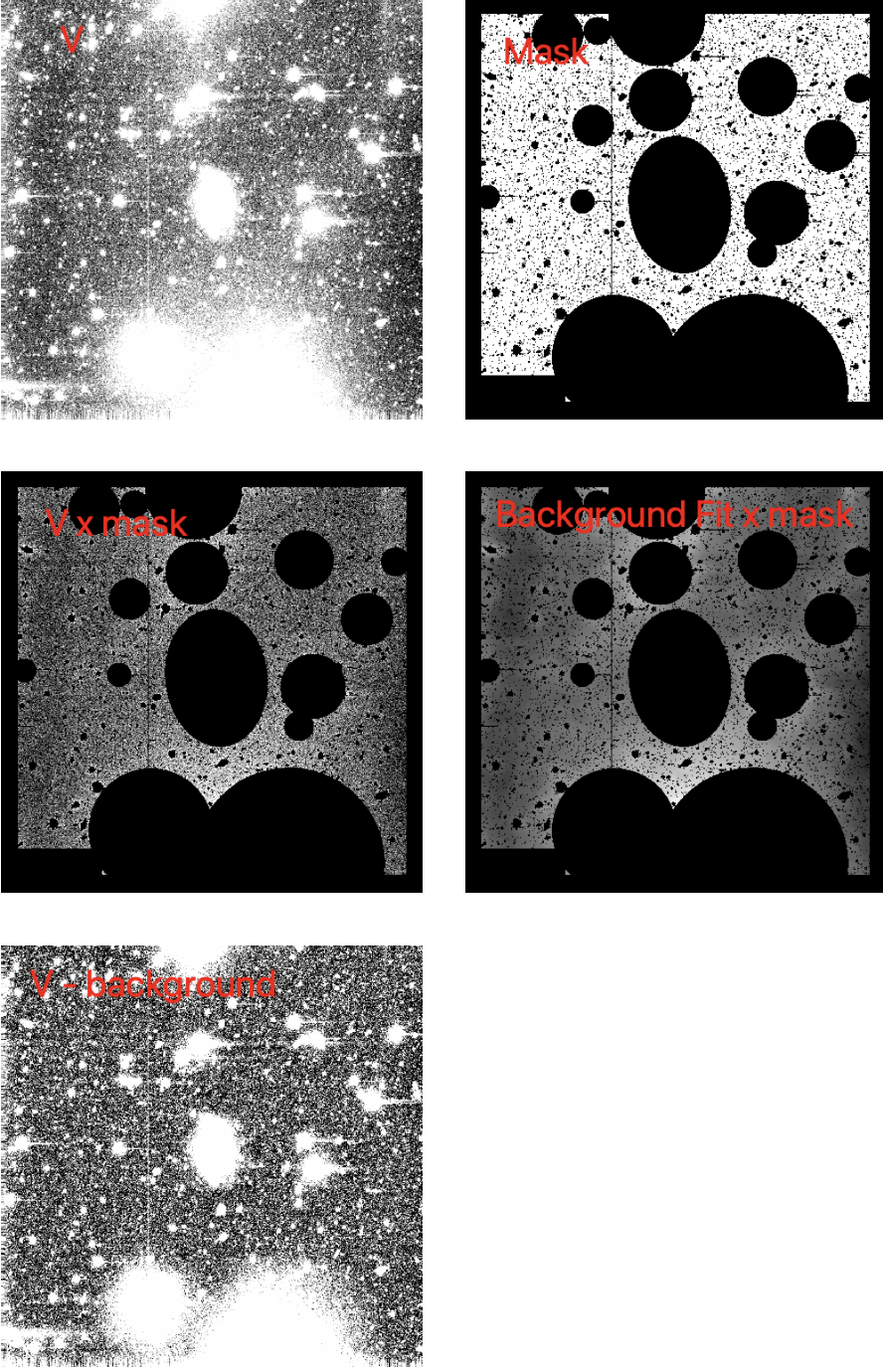}
\vskip -0.1truein
\caption{The process of sky/background fitting and subtraction is illustrated here for DDO 43 as an example.
Upper left: $V$ image of DDO43. North is up and East to the left. This is displayed with a linear stretch from counts 2415 to 2445.
Upper right: Mask of objects not to be included in the fit of the sky/background. This is displayed 0 to 1, 
where 0 are the pixels not to be fit.
Middle left: $V$ image times the mask. The mask is applied in the fitting, but this image is given to make clear what is 
being fit in this case. This is displayed with a stretch from counts 2415 to 2445.
Middle right: The fit to the sky/background times the mask. This is displayed with a stretch from counts 2415 to 2445
for direct comparison to the frame to the left.
In this case the fitting process used the python package {\sc Background2D} 
with a 13 pixel {\sc boxsize} and a 7 pixel {\sc filtersize}.
Bottom left: $V$ minus the fit to the sky/background. This is displayed with a stretch from counts -3 to 10.
\label{fig-back}}
\end{figure}

The sky/background in each galaxy image in each filter was fit and subtracted in the following manner. 
First, a mask was constructed to remove certain parts of the image from the fitting process. These included
the galaxy itself with a wide margin to allow for the faint far outer disk, bright stars and any associated saturation spikes,
fainter stars and background galaxies identified by Source Extractor, and the edges of the image that were 
affected by the shifting of individual images to align those that comprised the co-added final image.
Second, the unmasked part of the image was fit 
with a two-dimensional polynomial of various x and y orders and with the python package
{\sc Background2D} with various {\sc boxsize} and {\sc filtersize} parameters. We found the lowest order background fit
that cleanly produced a sky value at zero across the image. 
Finally, this sky/background image was subtracted from the original image.
We illustrate this process in Figure \ref{fig-back} for DDO 43 as an example.

\clearpage

\subsection{Analysis} \label{sec-analysis}

\begin{figure}[t!]
\epsscale{1.0}
\plotone{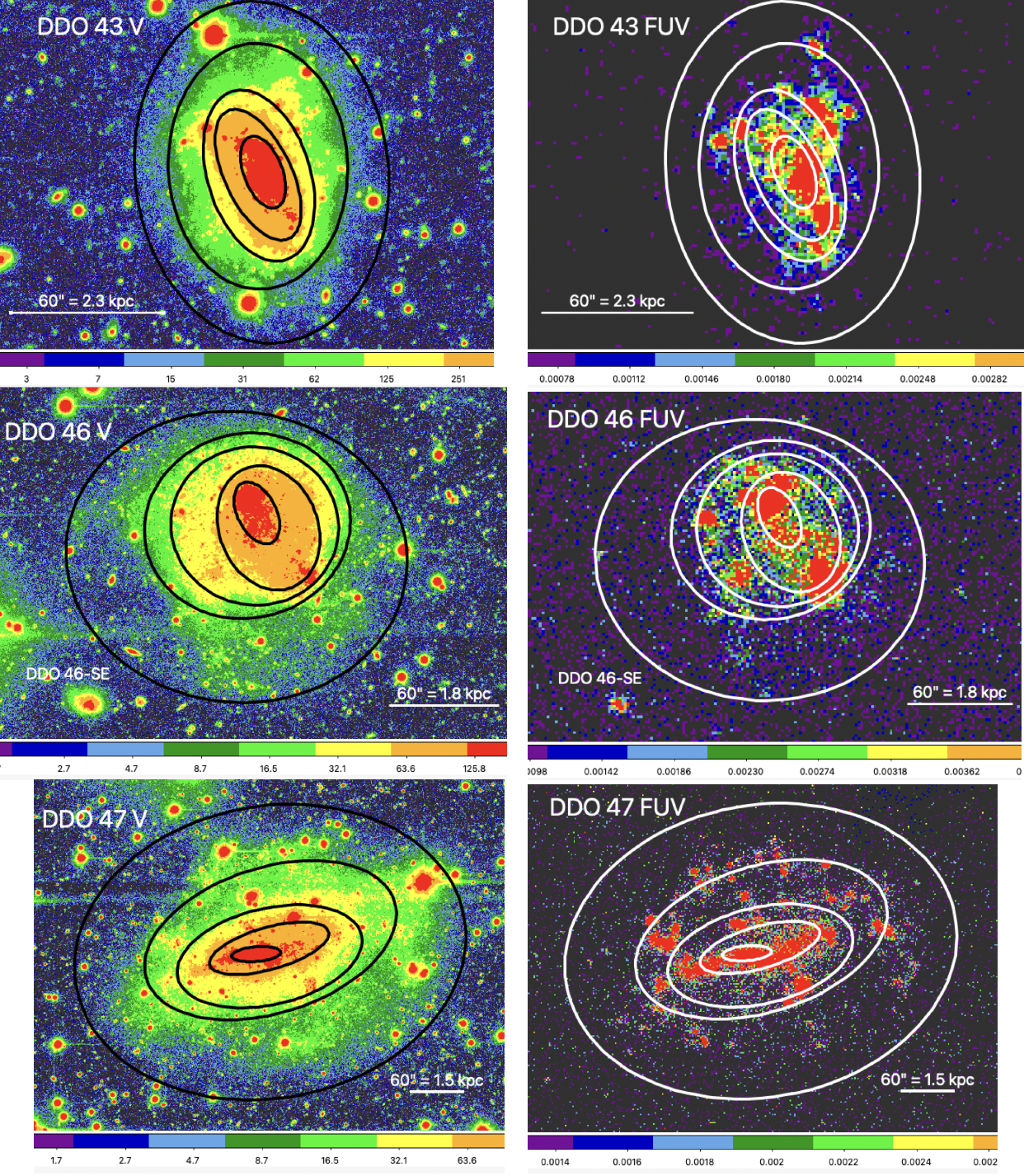}
\caption{$V$ images displayed in log scale and FUV images displayed in linear scale of each galaxy. North is up and East to the left.
The ellipses fit obvious isophotes revealed in the images by the colors.
The $V$ ellipses are plotted on the FUV image to compare the broad-band stellar disk structure
with the star formation activity.
The parameters of the ellipses are given in Table \ref{tabell}.
\label{fig-d43-d46-d47v_fuv}}
\end{figure}

\begin{figure}[t!]
\epsscale{1.0}
\plotone{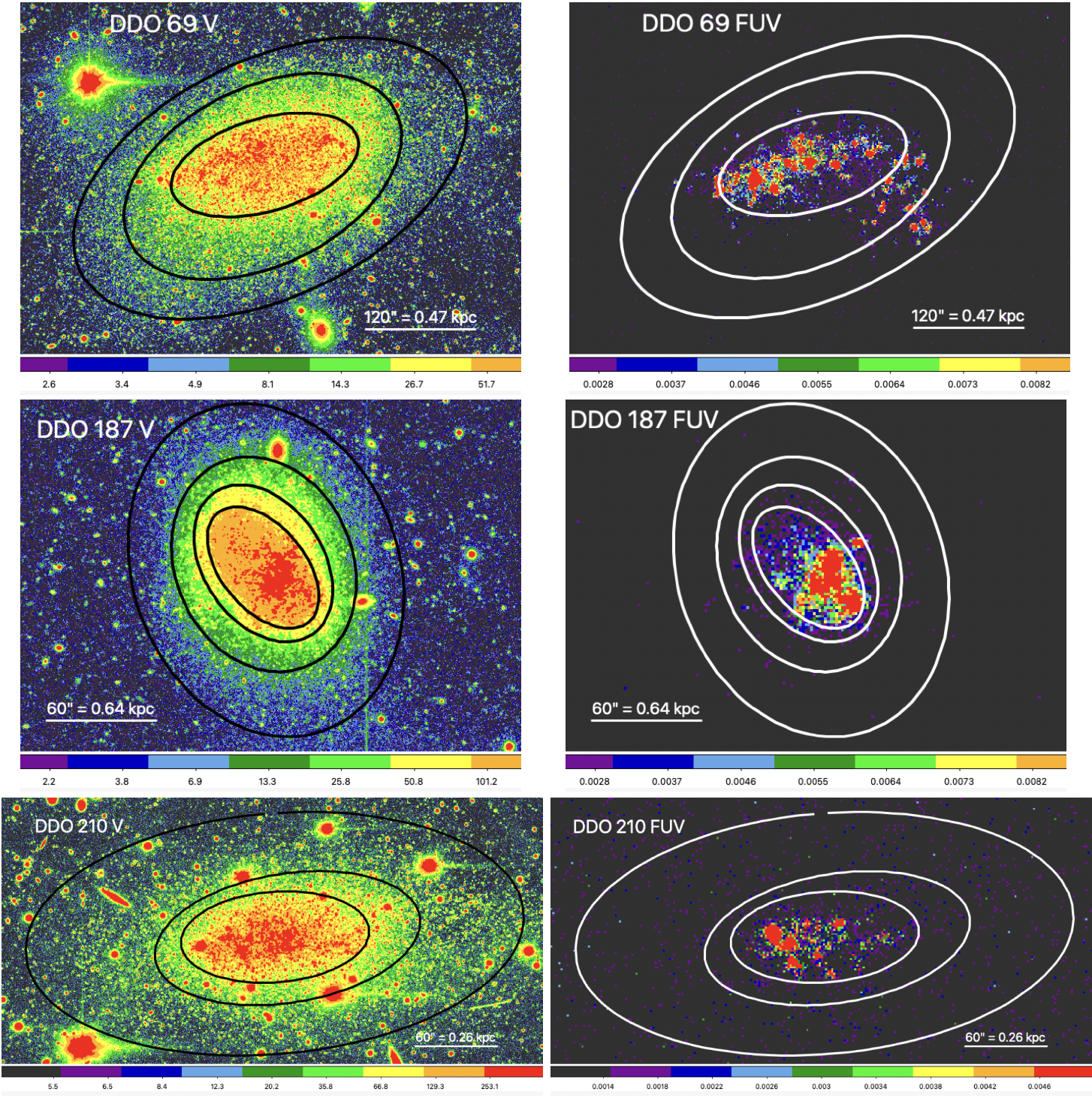}
\caption{As in Figure \ref{fig-d43-d46-d47v_fuv}.
\label{fig-d69-d187-d210v_fuv}}
\end{figure}

\begin{figure}[t!]
\epsscale{1.0}
\plotone{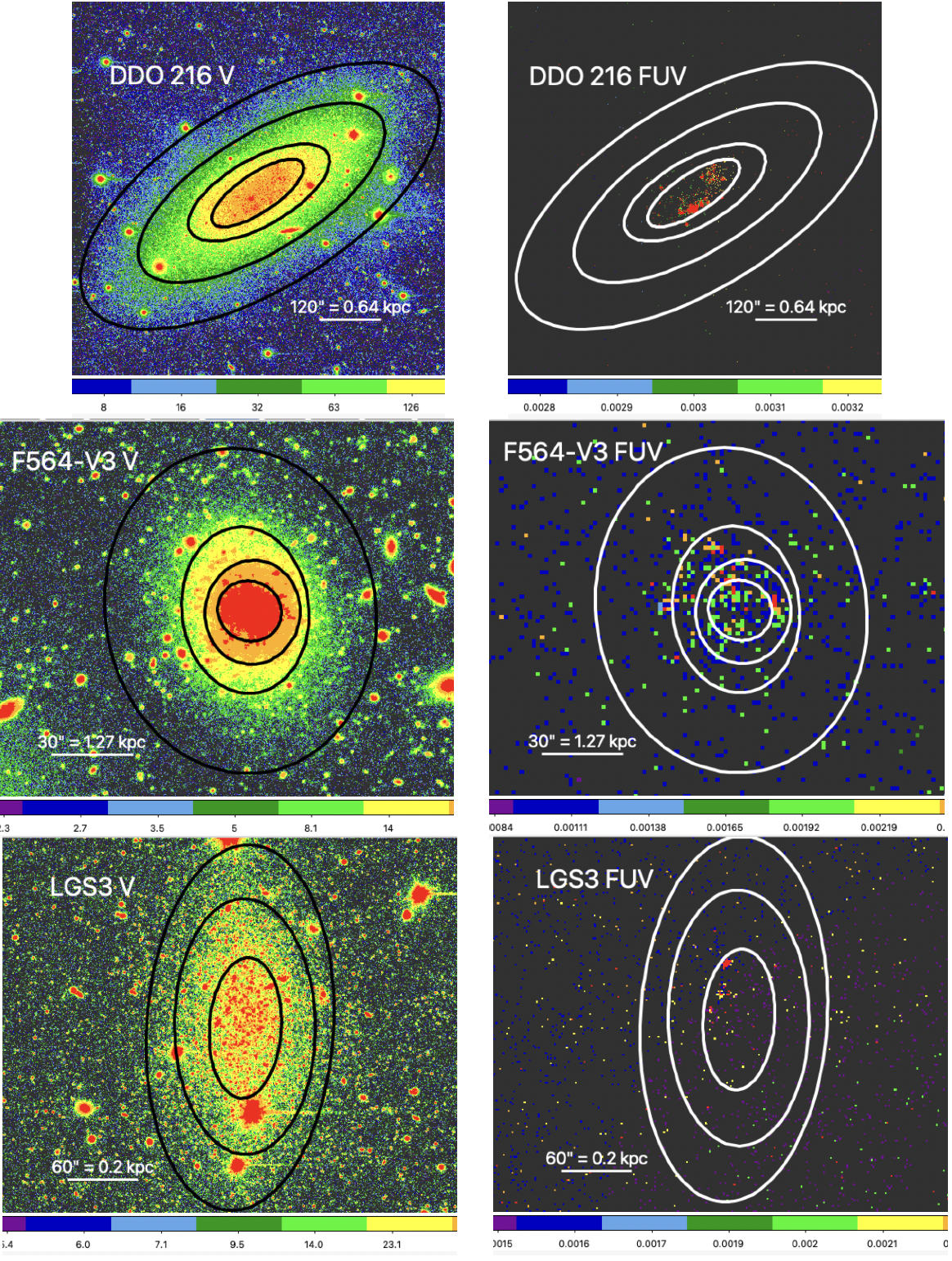}
\caption{As in Figure \ref{fig-d43-d46-d47v_fuv}.
\label{fig-d216-f564v3-lgs3v_fuv}}
\end{figure}

\begin{figure}[t!]
\epsscale{1.0}
\plotone{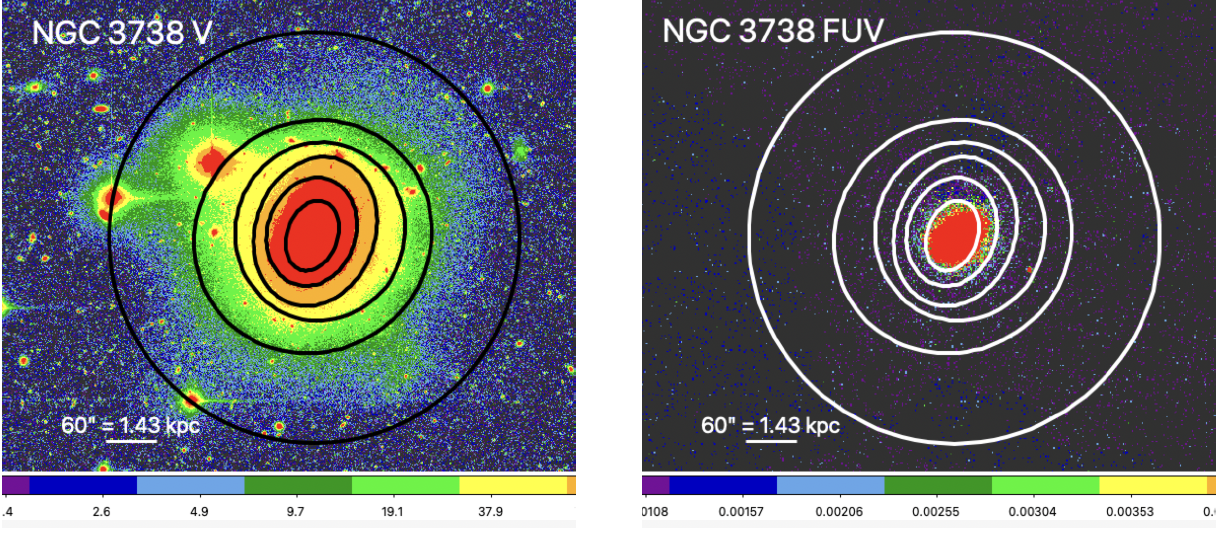}
\caption{As in Figure \ref{fig-d43-d46-d47v_fuv}.
\label{fig-n3738v_fuv}}
\end{figure}

The sky-subtracted $V$ and FUV images of each galaxy are shown in Figures \ref{fig-d43-d46-d47v_fuv}, \ref{fig-d69-d187-d210v_fuv},
\ref{fig-d216-f564v3-lgs3v_fuv}, and \ref{fig-n3738v_fuv}.
We display each galaxy's sky-subtracted $V$ image in a logarithmic scale to better see both inner and outer structures.
The images are displayed in color, where red is relatively high intensity and blue or purple are low intensities
(see color bars in the figures).
In these images, so displayed, one can immediately see different intensity levels (e.g. isophotes). 
For example, in DDO 43 in Figure \ref{fig-d43-d46-d47v_fuv}, one can see isophote levels that outline the 
red, orange, yellow, green, and blue regions. 
We targeted these brightness levels in our isophote fitting.
This allows us to probe disk structure at different extremes in the stellar disk, particularly the very low surface brightness far outer disk where star formation is proceeding very slowly and the inner disk where star formation is currently obvious.
This approach, rather than measuring the surface brightness at fixed radii, allows us to prioritize and characterize structural transitions
in the galaxy disks, if they are present.

To fit ellipses to the isophotes, we used python's {\sc photutils.isophote.ellipse} and {\sc scipy.optimize.brentq}.
Foreground stars and background galaxies were masked in the images prior to fitting.
For each isophote, we targeted a particular surface brightness level in counts per pixel and solved for the semi-major axis length $a$ 
at which the mean intensity along the ellipse approximately matched the target value.
For each isophote, an initial estimate of the ellipse parameters (center, ellipticity, position angle, and range of plausible $a$ values) 
was obtained from visual inspection but none of the parameters were fixed in the fitting.
The algorithm then iteratively refined the solution by adjusting the semi-major axis over the range of allowed values
while recomputing the mean intensity along an ellipse, following the method of \citet{Jedrzejewski87}. 
The semi-major axis corresponding to each target intensity was determined using the 
bracketing root-finding algorithm {\sc brentq} applied to the difference between the measured mean isophotal intensity and the target value.
The fitting algorithm samples the intensity along the ellipse and performs a harmonic expansion of the azimuthal light distribution, 
allowing refinement of the ellipse parameters. 
The resulting fits gives us the centers, PAs, $a$, and $b/a$ ratios of each isophote as a function of surface brightness.

In a few cases the fitting algorithm was unstable or unable to fit an ellipse to an isophote. One problem is with galaxies that 
are close enough to be resolved. The algorithm was unable to reasonably fit the inner isophote of these galaxies, so we
smoothed the images using boxcar smoothing and fit isophotes to the smoothed image. For DDO 69 we used a 50 $\times$ 50 pixel
smoothing window and for DDO 210 and LGS3 we used a 10 $\times$ 10 pixel smoothing window.
The other problem was in the very low surface brightness of the far outer isophote. This was a problem in most cases
for that isophote, so we used the 29 \optsb\ ellipses determined by \citet{hunter25}. 
They created contours of the specified surface brightness with a smoothness factor of 8, which enabled them to reach very low
count levels with a more robust method \citep[for details on their fitting process see][]{hunter25}. 
The exceptions were DDO 46, DDO 47, and DDO 216 in which the 29 \optsb\ isophote was included in the fits.

\begin{figure}[t]
\vskip 0.2truein
\plotone{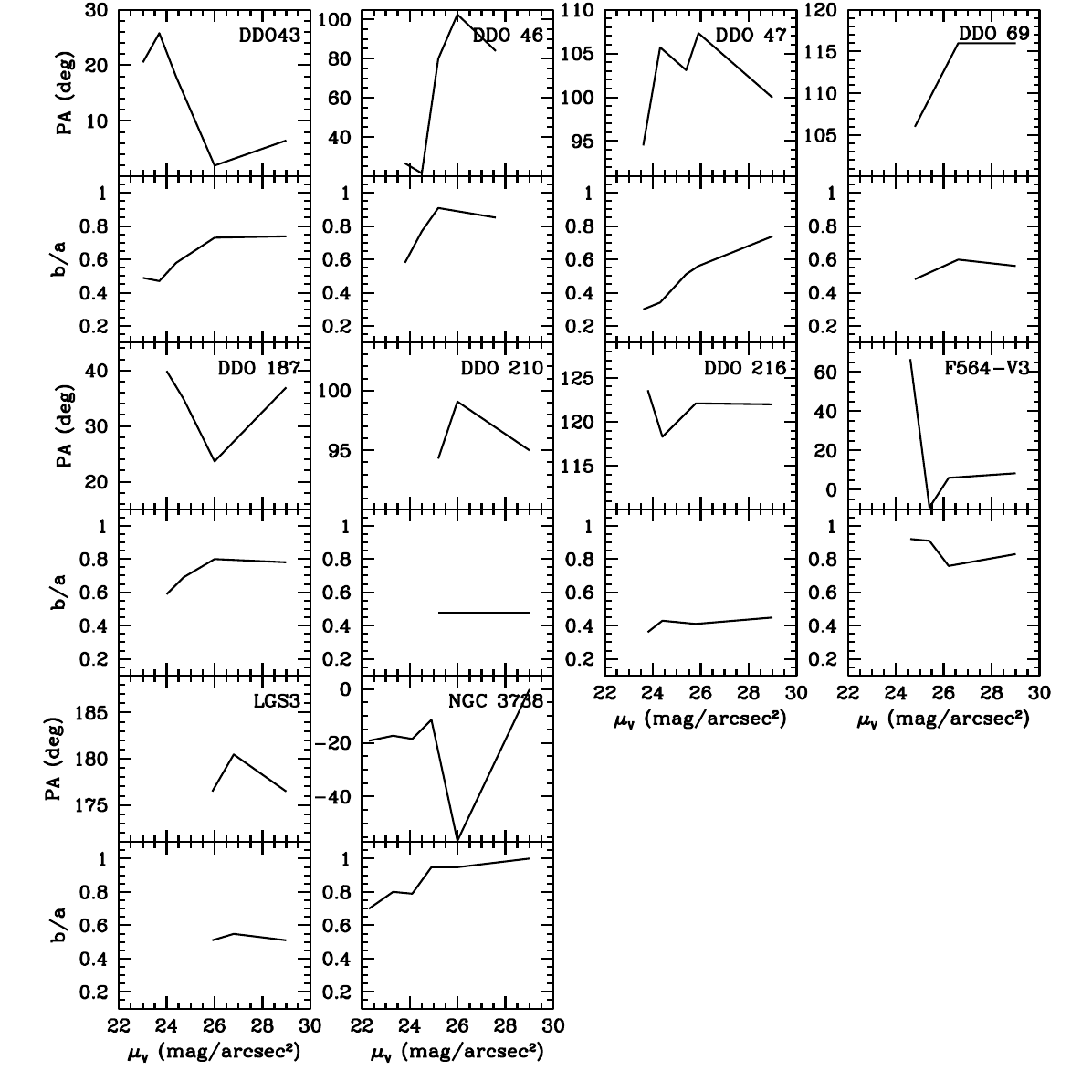}.   
\caption{
PA and $b/a$ of isophotes as a function of surface brightness in $V$ for each galaxy. 
The $b/a$ panel is below the PA panel for each galaxy.
}
\label{fig-pavsmuv}
\end{figure}

The ellipses defined on the $V$ image were drawn as is on the FUV image of each galaxy.
Figures \ref{fig-d43-d46-d47v_fuv}, \ref{fig-d69-d187-d210v_fuv},
\ref{fig-d216-f564v3-lgs3v_fuv}, and \ref{fig-n3738v_fuv} show the $V$ and FUV images with the isophote ellipses.
The ellipse parameters are given in Table \ref{tabell}
along with the radial offset ($R_\Delta$) of each ellipse from the center of the galaxy  \citep[centers from][]{he06} in the plane of the galaxy
and the difference between the PA of each inner ellipse with an outer ellipse ($\Delta$PA).
Ellipse PAs and $b/a$ are shown as a function of $V$ surface brightness for each galaxy in Figure \ref{fig-pavsmuv}.

\clearpage

\begin{deluxetable}{clccccccc}
\footnotesize
\tablecaption{Isophotoal Ellipse Parameters. \label{tabell}}
\tablewidth{0pt}
\tablehead{
\colhead{} & \colhead{} & \colhead{$a$\tablenotemark{b}} & \colhead{PA} & \colhead{} & \colhead{Center\tablenotemark{c}} & \colhead{$\mu_V$} & \colhead{$R_\Delta$\tablenotemark{d}} & \colhead{$\Delta$PA\tablenotemark{e}}\\
\colhead{Galaxy} & \colhead{ID\tablenotemark{a}} & \colhead{(arcsec)} & \colhead{(deg)} & \colhead{$b/a$} & \colhead{(arcsec,arcsec)} & \colhead{\optsb} & \colhead{(pc)} & \colhead{(deg)} \\
}
\startdata
DDO 43   & 1 &   14.8 &  20.5 & 0.49 &  -0.55   0.06 & 23.0 &   32 &  -14.0 \\
         & 2 &   27.8 &  25.8 & 0.47 &   1.39  -0.99 & 23.7 &   91 &  -19.3 \\
         & 3 &   34.6 &  17.9 & 0.58 &   1.06  -1.35 & 24.4 &   83 &  -11.4 \\
         & 4 &   48.2 &   2.0 & 0.73 &   1.45   2.33 & 26.0 &  115 &    4.5 \\
         & 5 &   67.1 &   6.5 & 0.74 &   0.00  -0.00 & 29.0 &    0 &    0.0 \\
DDO 46   & 1 &   18.1 &  26.7 & 0.58 &  -4.69   8.12 & 23.8 &  310 &   57.3 \\
         & 2 &   34.8 &  21.6 & 0.77 & -11.11  -0.00 & 24.5 &  329 &   62.4 \\
         & 3 &   45.8 &  80.1 & 0.91 &  -3.56   1.03 & 25.2 &  112 &    3.9 \\
         & 4 &   56.5 & 102.6 & 0.89 &   0.44   1.14 & 26.0 &   40 &  -18.6 \\
         & 5 &   93.2 &  84.0 & 0.85 &   6.41 -15.82 & 27.6 &  570 &    0.0 \\
DDO 47   & 1 &   26.9 &  94.5 & 0.30 &  15.19  -5.40 & 23.6 &  430 &    5.5 \\
         & 2 &   67.3 & 105.7 & 0.34 &   1.04   0.87 & 24.3 &   64 &   -5.7 \\
         & 3 &  102.8 & 103.1 & 0.51 &   0.54  -7.90 & 25.4 &  451 &   -3.1 \\
         & 4 &  141.7 & 107.3 & 0.56 &  -0.52   9.76 & 25.9 &  559 &   -7.3 \\
         & 5 &  215.0 & 100.0 & 0.74 &  -0.01  -3.20 & 29.0 &  185 &    0.0 \\
DDO 69   & 1 &  104.3 & 106.0 & 0.48 &   5.53  14.50 & 24.8 &  121 &   10.0 \\
         & 2 &  159.9 & 116.0 & 0.60 &   8.09   2.04 & 26.6 &   49 &    0.0 \\
         & 3 &  228.0 & 116.0 & 0.56 &   0.00  -0.00 & 29.0 &    0 &    0.0 \\
DDO 187  & 1 &   38.3 &  39.9 & 0.59 &   1.52   1.44 & 24.0 &   23 &   -2.9 \\
         & 2 &   46.5 &  34.9 & 0.69 &   1.36   3.71 & 24.7 &   43 &    2.1 \\
         & 3 &   59.7 &  23.7 & 0.80 &   1.07   3.21 & 26.0 &   37 &   13.3 \\
         & 4 &   92.0 &  37.0 & 0.78 &   0.00  -0.00 & 29.0 &    0 &    0.0 \\
DDO 210  & 1 &   67.9 &  94.3 & 0.48 &  -0.20  -2.34 & 25.2 &   26 &    0.7 \\
         & 2 &   96.4 &  99.1 & 0.48 &  -9.18  -3.16 & 26.0 &   59 &   -4.1 \\
         & 3 &  180.0 &  95.0 & 0.48 &   0.00  -0.00 & 29.0 &    0 &    0.0 \\
DDO 216  & 1 &  105.2 & 123.6 & 0.36 &   4.83   5.65 & 23.8 &  112 &   -1.6 \\
         & 2 &  153.8 & 118.3 & 0.43 &   2.71   5.94 & 24.4 &   98 &    3.7 \\
         & 3 &   36.6 & 122.1 & 0.41 &  -2.86   6.13 & 25.8 &   64 &   -0.1 \\
         & 4 &  160.0 & 122.0 & 0.45 &   0.00  -0.00 & 29.0 &    0 &    0.0 \\
F564-V3  & 1 &   11.9 &  66.7 & 0.92 &  -3.33  -0.06 & 24.6 &  173 &  -58.2 \\
         & 2 &   19.2 &  -9.0 & 0.91 &  -4.43  -0.54 & 25.4 &  229 &   17.5 \\
         & 3 &   30.7 &   6.0 & 0.76 &  -1.78   0.27 & 26.2 &   94 &    2.5 \\
         & 4 &   60.0 &   8.5 & 0.83 &   0.00  -0.00 & 29.0 &    0 &    0.0 \\
LGS3     & 1 &   59.9 & 176.5 & 0.51 &  -4.42  -0.27 & 25.9 &   35 &    0.0 \\
         & 2 &  108.9 & 180.6 & 0.55 &  -4.13   0.92 & 26.8 &   32 &   -4.1 \\
         & 3 &  156.0 & 176.5 & 0.51 &   0.00  -0.00 & 29.0 &    0 &    0.0 \\
NGC 3738 & 1 &   42.0 & -19.3 & 0.70 &   2.40   2.54 & 22.3 &   83 &   19.3 \\
         & 2 &   64.4 & -17.4 & 0.80 &   3.18   7.12 & 23.3 &  185 &   17.4 \\
         & 3 &   88.6 & -18.5 & 0.79 &  -1.13   7.52 & 24.1 &  181 &   18.5 \\
         & 4 &  104.2 & -11.4 & 0.95 &  -6.78   7.52 & 24.9 &  241 &   11.4 \\
         & 5 &  141.0 & -56.4 & 0.95 &   1.56   1.21 & 26.0 &   47 &   56.4 \\
         & 6 &    0.0 &   0.0 & 1.00 &   0.00  -0.00 & 29.0 &    0 &    0.0 \\
\enddata
\tablenotetext{a}{Ellipse No.\ 1 is the innermost ellipse.}
\tablenotetext{b}{Semi-major axis, $a$, of the ellipse in arcseconds.}
\tablenotetext{c}{Offset of the ellipse center from the center of the galaxy 
as given in Table 1, in arcseconds: E-W with positive E, N-S.}
\tablenotetext{d}{Radial offset, $R_\Delta$, of the ellipse center from the galaxy center 
in parsecs in the plane of the galaxy. 
Galaxy inclinations are given by \citet{lt}.}
\tablenotetext{e}{Difference in PA, $\Delta$PA, of the ellipse relative to that of an outer ellipse. The sense is PA of outer ellipse minus PA of the other ellipses.}
\end{deluxetable}

\clearpage

\clearpage
 
We have converted our elliptical isophotes for each galaxy into a model galaxy and subtracted that from the observed
$V$ image. 
The inputs to the model are the parameters of the ellipses fit to isophotes.
The model then interpolates log-intensity between isophotes. 
The PAs and ellipticities of the ellipses are interpolated linearly, producing 
a smooth PA twist between ellipses. 
Note that our elliptical isophotes are aimed at identifying large-scale structures, so the model cannot contain smaller structures or
deviations from elliptical structures. The original $V$ image (minus sky), the model, and the $V$ minus model image
are shown in Figures \ref{fig-models1}, \ref{fig-models2}, and \ref{fig-models3} in Appendix \ref{models}.

In a few instances we measured the colors of interesting pieces of the far outer disk of individual galaxies in order to estimate
the age of the stellar population there. We also measured the colors in annuli of the stacked image discussed below.
To estimate ages, we used the complex stellar population model of \citet{bc03}
with an exponentially declining SFR for a typical metallicity of $Z=0.004$. 
We fit $U-B$, $B-V$, and $V-I$ as a function of $\log$ age from the model with an order three polynomial
and used the fits to determine the age and uncertainty for each color available for the region.
We then determined an age for each region from the average of the ages from the different filter combinations.

\section{Discussion} \label{sec-discuss}

If the ellipses fitting the different $V$ surface brightness contours of the \dirr\ galaxies were well behaved, having 
the same PA, $b/a$, and centers, we would have nothing further to say. However, that is not what we found.
Many of the galaxies in our sample have ellipses that are different in these parameters as one proceeds from higher surface brightness
in the central regions to the much lower surface brightnesses in the far outer stellar disk.
This led us to superpose the $V$-band ellipses on the FUV images, which trace recent star formation.
Here we discuss the effects we found of the current star formation on the apparent structure of the disks,
even in the broad-band optical.
Finally, we address the possibility of the signatures of stellar haloes.  

\subsection{Star formation and the structure of stellar disks} \label{sec-sf}

We see that the inner 1-3 ellipses of 8 of the 10 galaxies have significant differences in PA, center, and/or $b/a$
compared to those of the outer ellipses. 
Comparison with their placement on the galaxies' FUV images, which trace recent star formation, shows 
that these changes coincide with the location of most of the FUV emission.
The significant non-zero differences of PAs range from 3\arcdeg\ to 100\arcdeg, 
offsets of the centers range from  23 to 570 pc,  
and $b/a$ differ up to 0.4. 
NGC 3738, like other BCDs, has intense star formation activity that is centrally concentrated, and it has the
most extreme change in isophotes from the central region to the outer parts,
being elliptical in the central regions and circular in the outer regions.

It makes sense that star formation in low surface brightness systems might skew the isophotes,
making patches of the galaxy significantly brighter for a relatively short time
(see the FUV images in Figures \ref{fig-d43-d46-d47v_fuv}, \ref{fig-d69-d187-d210v_fuv}, \ref{fig-d216-f564v3-lgs3v_fuv},
and \ref{fig-n3738v_fuv}).
For example, the inner two ellipses in DDO 43  
are skewed by two off-center FUV complexes that
are 400 pc and 250 pc in diameter and in DDO 46 by complexes that are 440 pc and 350 pc in diameter.
In DDO 47 the inner ellipse is affected by one large off-center complex with a diameter of 580 pc,
and that complex is part of a long thin FUV structure that is 2.2 kpc long and affects the second and third ellipses.
The inner ellipse in DDO 69 is dominated by one long off-center complex that is 370 pc long.
And the inner ellipse of DDO 216 is defined by two FUV regions each of order 140 pc in diameter.
The sizes of these FUV complexes represent 0.3 to 1.9 $R_D$, and thus are large for such tiny galaxies.

However, what is notable in these galaxies is that
the changes found in the $V$ isophotes extend to low surface brightness levels: ranging from 
23.7 \optsb\ to 25.4 with the rest around 24.9 \optsb. Thus, patchy star formation occurs even in the outer disk
and not just in the central regions or brightest parts of the galaxies.

There are only two galaxies with little change in ellipse parameters with radius: DDO 216 and LGS3. 
In both cases the FUV emission is contained within the inner ellipse. 
In the case of LGS3 there is very little star formation going on even in the center,
and the inner ellipse outlines a low $V$ surface brightness of 25.9 \optsb. 
On the other hand, F564-V3 also has very little FUV emission but its inner two ellipses are offset relative to the
outer ellipses by 173 pc and 229 pc and the innermost ellipse also has a different PA by 58\arcdeg.
DDO 216 has somewhat more FUV emission than LGS3. Even so, its inner ellipses are only offset from the outer ellipses
by 112 pc and 98 pc and a few degrees in PA.

We conclude that the star formation activity in these \dirr\ galaxies skews the broad-band
optical isophotes.
The stellar disks are low surface brightness and asymmetries in the distribution of the star formation 
are bigger relative to the size of the galaxy than in giant spiral galaxies.

\subsection{External influences}  \label{sec-external}

The galaxies in the original sample of \dirr\ galaxies that 
the ultra-deep imaging sample was derived from were chosen not to be 
obviously interacting with another galaxy.
However, three of our 10 galaxies show evidence that some past  merger has probably taken place recently
enough that we still see significant anomalies introduced by the event.

\subsubsection{DDO 46}

DDO 46 is the most extreme case in terms of asymmetry of isophotes. It has an extension of the stellar disk to the south
that required \citet{hunter25} to offset the $V$ 29 \optsb\ ellipse 570 pc, mostly south, relative to the center of the 26 \optsb\ ellipse.
They found several distinct stellar associations in the southern extension of DDO 46 that coincide with FUV emission
and have young ages. 
One object has colors consistent with an age of $\sim$30 Myr, and two others have ages $\sim$30-100 Myr.
The underlying disk within which these associations are located is relatively old (about 1 Gyr).
We also see that there is a small protrusion of the disk to the east; 
its age is also about a Gyr. 

However, most notable, given the southern extension, is a detached object that looks like a tiny
galaxy 2.2\arcmin\ to the southeast  (7:41:34.22, 40:05:04) of DDO 46's center. 
We will refer to this as DDO46-SE (see Figure \ref{fig-d43-d46-d47v_fuv}).
DDO46-SE is at approximately the same distance as DDO 46 because it is detected in \HI-line channel maps (308.1 - 328.7 \kms)
overlapping with DDO 46's velocities
\citep[][see their Figure 15]{lt}. 
(We note that there is another resolved source 28\arcsec\ to the west of DDO46-SE, but it is not seen in the \HI\ map
of DDO 46 and there are many other background galaxies around. Therefore, we conclude that this source is probably
background to DDO 46.)
DDO46-SE does not show up on its own in any catalog of galaxies.
From an observed $V$ magnitude of $18.62 \pm 0.004$,
we find a luminosity for DDO46-SE of ($1.15\pm0.005) \times 10^6$ L\solar.
Using a stellar mass-to-light ratio $M_*/L_V$ of $0.517 \pm 0.006$ deduced from the 
formulation of $M_*/L_V$ as a function of $B-V$ for \dirr\ galaxies by \citet{kim16},
we find a stellar mass $M_*$ of ($5.9\pm0.1) \times 10^5$ M\solar.
DDO46-SE's colors are consistent with a moderate age of 800 Myr and it is also visible in the FUV image
of DDO 46.
DDO 46 itself has an
$L_V$ of $(6.4 \pm 0.09) \times 10^7$ L\solar\ and $M_*$ of $(3.0 \pm 0.09) \times 10^7$
\citep{he06}.
Thus, DDO46 has 50 times the stellar mass of DDO46-SE.

\begin{figure}[t!]
\epsscale{0.4}
\vskip -0.8truein
\plotone{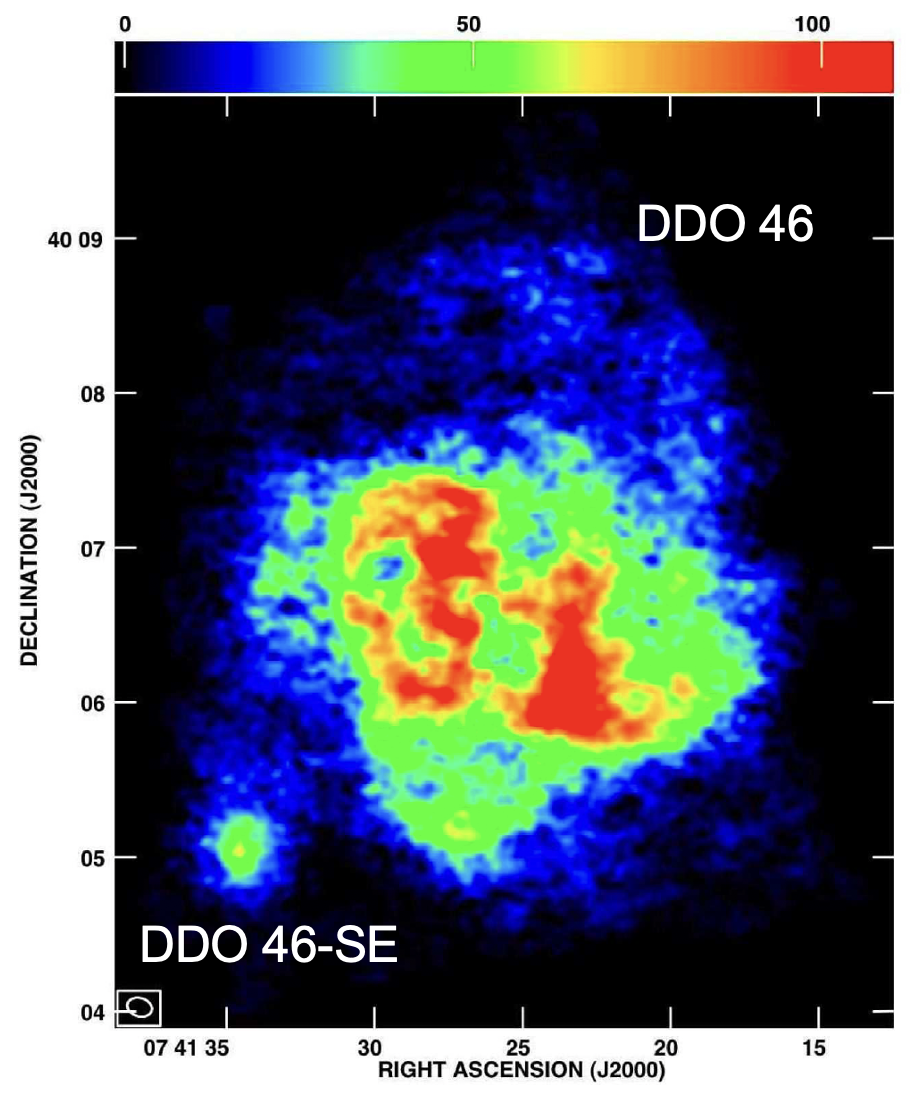}
\caption{Reproduction of the integrated \HI\ (moment 0) map of DDO 46 made from the Hanning-smoothed Naturally-weighted cube
from \citet{lt}.
\label{fig-d46mom0}}
\end{figure}

It is tempting to think that DDO46-SE is responsible for the peculiar southeast extension of DDO 46
through a galaxy-galaxy interaction, especially given the galaxy-like morphology of DDO46-SE.
And in fact there is a bridge of \HI\ extending north from DDO46-SE and connecting with the east side
of DDO 46 \citep[Figure 16 in][reproduced here in Figure \ref{fig-d46mom0}]{lt}. 
DDO46-SE appears otherwise detached from DDO 46 itself both in the optical and in \HI.
However, there is also an off-center extension of \HI\ to the north of DDO 46, and a peculiar extension of the rotation curve
to the west on the south edge of the \HI\ distribution.
But, there are other characteristics that are not consistent with an interaction scenario:
1) The \HI\ velocity of DDO46-SE matches that of the east side of DDO46.
Also, the southern extension of DDO 46 seen in the optical is a part of the overall
\HI\ morphology and kinematics of DDO 46 \citep{lt}.
2) There is nothing unusual in the strong and weak non-circular
gas motions identified by \citet{oh15}. 
Furthermore, the PA deduced from the velocity field of the gas is close to (within 10\arcdeg) of
the PA of the optical galaxy seen in $V$ images.
3) There is also nothing unusual evident in the gas velocity dispersion of either object \citep{lt}.
4) Finally, DDO46-SE does not look disturbed optically, while DDO 46 has the large peculiar southern extension.
Could a mass ratio of 50, nevertheless, allow the smaller DDO46-SE to disturb the much larger DDO 46 to that extent
without itself being messed up?
Whether DDO46-SE is responsible for the southern extension of DDO 46 or not, however,
DDO 46 does have an unusual optical morphology that is hard to explain other than through an external 
interaction of some kind in the past.

\subsubsection{DDO 187}

The star formation in the central region of DDO 187 was studied by \citet{Gallardo18}.
Here, the area that they examined is within our ellipse No.\ 1. 
They were struck by the asymmetry in the location of the star formation:
most of the star formation in DDO 187 is located in one half of the region of ellipse No.\ 1. 
This is easily seen in the FUV image of DDO 187 in Figure \ref{fig-d69-d187-d210v_fuv}.
The southwest side of ellipse No.\ 1 is the high star formation (HSF) side of \citet{Gallardo18}
and the northeast side of ellipse No.\ 1 is their low star formation (LSF) side.
They found that the HSF side has 30\% to 70\% higher pressure and gas density
compared to the LSF side.
At both the northeast and southwest ends of ellipse No.\ 1 are regions of high \HI\ velocity dispersion (14-20 \kms)
and complex kinematics.
\citet{Gallardo18} argue that there must have been a merger affecting the galaxy.

\subsubsection{NGC 3738}

\citet{Gallardo18} also found a similar situation in NGC 3738. There too they found that the star formation
activity within our ellipse No.\ 1 is divided into HSF (northwest) and LSF (southeast) halves.
There are high \HI\ velocity dispersions (24-35 \kms) and complex and non-circular velocities 
to the northeast of this ellipse, as well as to the west. 
Ellipse No.\ 1 is located at a bend in the backward S-shaped iso-velocity contour.
In addition, \citet{Ashley17}  performed an extensive dissection of NGC 3738's \HI\ velocity field
and concluded that this BCD is morphologically and kinematically disturbed
with unusually concentrated \HI.  If this is the result of a merger, the merger would have to be
advanced because there are no tidal tails. However, \citet{Ashley17} could not rule out
ram pressure stripping as the mechanism for NGC 3738's unusual gas morphology and kinematics.

 \subsection{Search for stellar haloes} \label{sec-halo}

We looked for galaxies in which there is a change in PA or $b/a$ in the lowest surface brightness regions of the galaxies.
This could indicate a transition from a disk to a halo stellar population.
Figure \ref{fig-pavsmuv} shows the PA and $b/a$ as a function of disk surface brightness in the $V$ filter for each
galaxy in our sample. We then extend our data to lower surface brightness by stacking galaxies into one
master image in order to search for a stellar halo as a general feature of \dirr\ galaxies.

\subsubsection{Changes in PA and $b/a$}

DDO 47 shows a steady increase in $b/a$ from the inner ellipse to the outer most ellipse with only small
changes in PA.
The age of several regions in the far outer disk is several Gyr.
So the stars in these regions are relatively old but not ancient. 
DDO 43 also shows significant changes in PA and $b/a$ between the inner disk ellipses and outer disk ellipses.
Given these changes with radius, we cannot rule out an extended halo in these two galaxies.

For DDO 187, the PA of the  major axis increase steadily from the inner ellipse to the outer ellipse,
24 \optsb\ to 29 \optsb, the inner two ellipses being affected  by star formation. 
However, $b/a$ is nearly the same in the outer two ellipses, although higher than for the inner two ellipses.
Regions out at the edges of the 29 \optsb\  ellipse are old at ages of 500 Myr to 8 Gyr.  
\citet{d187} used color magnitude diagrams (CMD) to trace the old extended stellar disk.
They suggest that a halo $+$ disk structure is plausible in DDO 187. They traced the galaxy to 1.3 kpc radius,
and their data suggest that the halo
starts around the same radius as our furthest out ellipse ($\sim$1 kpc).


In DDO 69  we see a large change in PA, but not in $b/a$, in going from a $\mu_V$ of 25 \optsb\ to 27 \optsb\
and no change in going from 27 \optsb\ to 29 \optsb.
However, in Section \ref{sec-sf} we showed that the innermost ellipse outlines the star formation
and argued that, like in many of the other \dirr\ galaxies, the characteristics of the inner isophote
are skewed by the star formation activity in an otherwise very low surface brightness system. 
However, the far outer stellar population of DDO 69, also known as Leo A, 
has been studied by \citet{LeoA04}. 
They used red supergiant (RGB) stars to trace the stellar population of DDO 69 to even larger radii
than we have here.
They found an exponential disk to 1.3 kpc and a halo that becomes prominent from there outward to 1.9 kpc
where the RGB stars cut off. They trace an ellipse in the outer halo with a PA of 114\arcdeg.
\citet{LeoA18} also used CMDs to trace an old, metal-poor stellar halo from 1.7 kpc to 2.3 kpc. 
The change in the PA from our inner ellipse to our middle
ellipse lies between 0.2 kpc and 0.7 kpc radius, closer to the center than the 1.3 kpc disk edge claimed by \citet{LeoA04}.
However, the PA of the far outer ellipse determined by \citet{LeoA04} is 114\arcdeg, which is essentially the same as our
PA for the 27 \optsb\ and 29 \optsb\ ellipses (116\arcdeg).
We measured the colors of a region along the west side of the galaxy 
and another along the southeast side of the galaxy.
These regions have ages of order 2-5 Gyr.
Thus, it is possible that our far disk annulus is the beginning of the stellar halo found by \citet{LeoA04} and \citet{LeoA18}.


\subsubsection{Stacking}

To explore the presence of haloes further, we have stacked the images of 7 of the 10 ultra-deep galaxies 
(DDO 43, DDO 47, DDO 69, DDO 187, DDO 216, F564-V3, LGS3). The three that were not included 
either have very bright stars close to the galaxy image (DDO 210, NGC 3738) 
or have a peculiar far outer morphology (DDO 46).
 Stacking has been shown to reveal the faint outer parts of typical galaxies within a limited mass
range or morphological class, including the average brightness and color profiles \citep[see also][]{zheng15}.

To stack the images for each filter, we followed the following steps:
1)  Shift the centers of the galaxies to the same pixel values.
2)  Rotate each image to put the major axis along the x axis. The major axis that was used
was that of the ellipse fits to the far outer disk of the $V$ image.
3)  Deproject each image to face-on, using the optical minor-to-major axis ratio $b/a$ determined from fitting the outer disk with ellipses
and conserving flux.
4)  Change the pixel scales so that every galaxy has the same disk scale length $R_D$ in pixels. We used DDO 69 as the fiducial 
and the $R_D$ of $0.19\pm0.01$ kpc for DDO 69 determined by \citet{kim13} for a distance of 0.8 Mpc.
5)  Create a mask for each galaxy of the pixels to not include in the final summing. The masks consisted of 3 parts: a) the area beyond the ellipse of the galaxy with a large margin to allow for detection of the galaxy much further out than 29 \optsb, b) objects found by Source Extractor, c) saturated stars and trails. We then subjected the masks to the same transformations as the images: shifting, rotating, deprojecting, and changing pixel scale. Because Source Extractor identified some star formation parts of some galaxies in the central regions resulting
in those parts being masked and because the central regions are generally irregular due to star formation, 
we limited the surface photometry profiles of the stacked image to no further interior than 26 \optsb.
6) Sum the images using the Image Reduction and Analysis Facility's (IRAF\footnote[3]{IRAF was distributed by the 
 National Optical Astronomy Observatory, which was operated by the Association of Universities for Research
 in Astronomy (AURA) under cooperative agreement with the National Science Foundation.})
{\sc imcombine} task, applying the mask for each galaxy and clipping the third pixel below the median in each ordered set of pixels.
We then divide by the number of galaxies to produce an average image.

Because the stacked images are so deep, the photon noise is small, generally smaller than the points in plots shown below. 
Systematic uncertainties from the processes of sky subtraction and stacking are expected to dominate photon noise.
To estimate the systematic uncertainties we used a ``jackknife'' resampling of the galaxy stack.
This involved 7 tests of cycling through the 7 galaxies, 
eliminating one of the galaxies, recombining to form a new stacked image, measuring the circular photometry,
and producing the surface brightness profile. 
The uncertainty of the data points is then estimated from $\sigma^2 = 1/(N-1) \times \sum (x_i - \bar{x})^2$,
where $N$ is 7, $x_i$ are the surface brightness results for a given annulus from each of the 7 tests, and $\bar{x}$ is the average 
of the 7 tests. This process is done for each annulus. 
The jackknife uncertainties include photon noise from the individual galaxy images, and therefore we did not explicitly add Poisson noise. 
This process was also followed to determine uncertainties in the colors.

\begin{figure}[t!]
\epsscale{0.5}
\vskip -0.8truein
\plotone{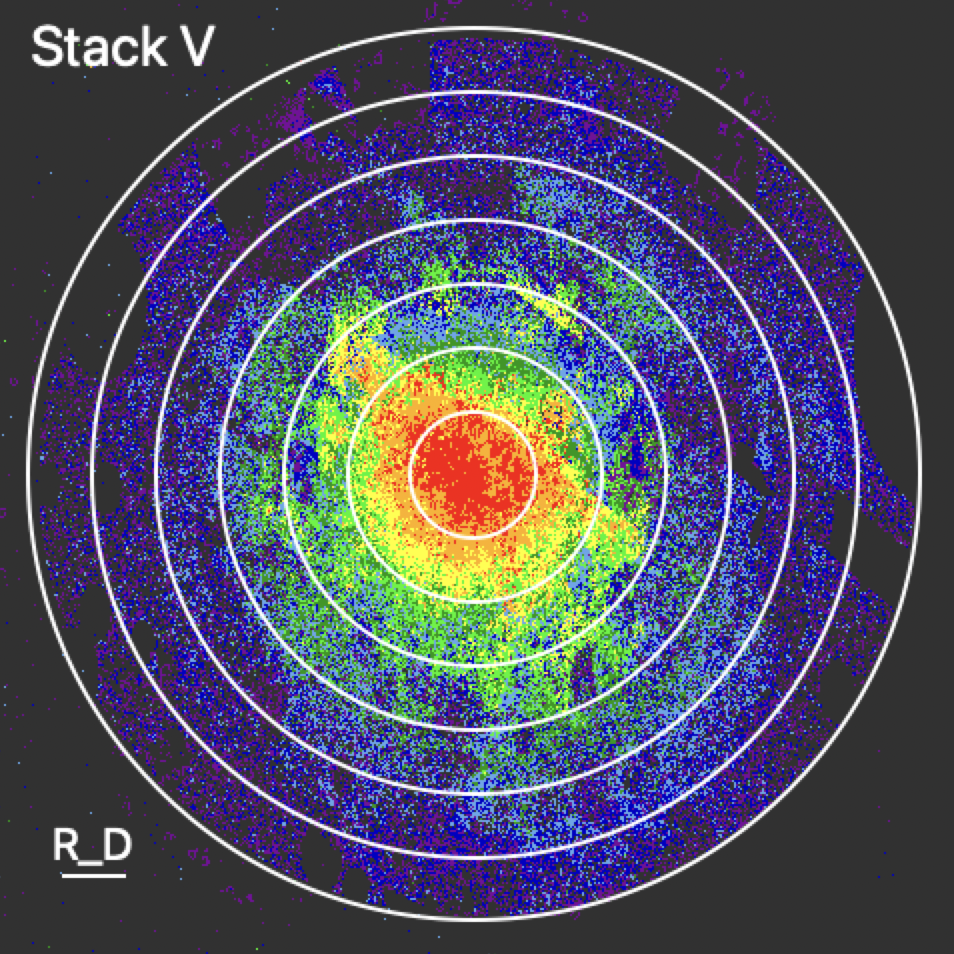}
\caption{Average stacked $V$ image produced by summing seven of the galaxies in our ultra-deep sample and dividing by the number
of galaxies. 
The circles are in radial steps of $R_D$ from 1 $R_D$ for the radius of the inner circle.
\label{fig-stackv}}
\end{figure}

\begin{figure}[h!]
\epsscale{0.8}
\plotone{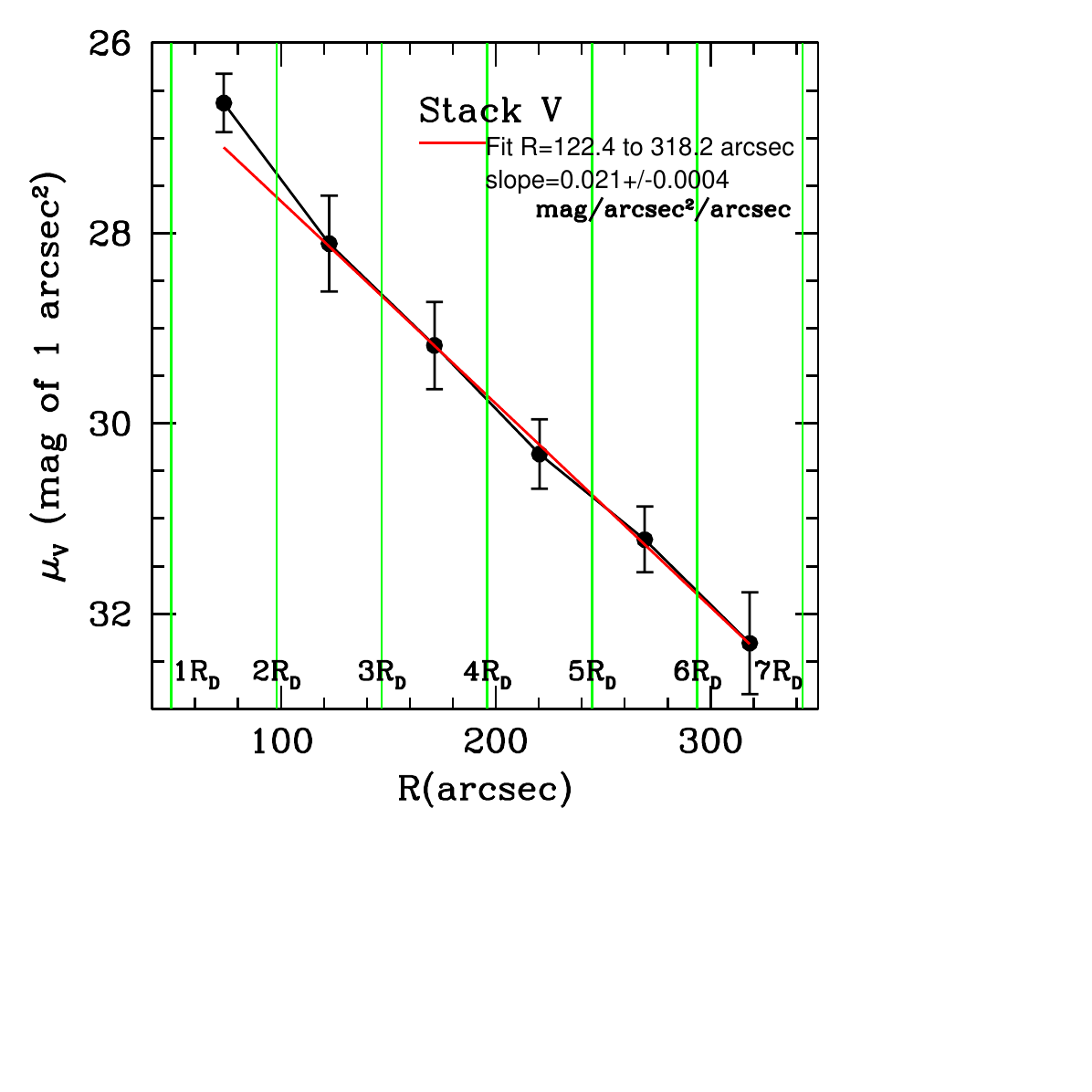}
\vskip -1.5truein
\caption{Surface photometry of the stacked $V$ image. 
The uncertainties are from a ''jackknife'' process described in the text.
The green vertical lines mark the radii of the circles in units of the $R_D$ of the image.
The surface brightness for an annulus is the average over the entire area of the annulus and is plotted
as a black filled circle at the mid-point of the annulus.
The red line is a linear fit to the second through sixth points and shows an exponential drop-off.
\label{fig-stackvprofile}}
\end{figure}

The stacked average $V$ image is shown in Figure \ref{fig-stackv}.
The circles are in radial steps of $R_D$ (49.0\arcsec).
Figure \ref{fig-stackvprofile} shows the surface brightness profile as a function of radius measured in the annuli between the circles shown in 
Figure \ref{fig-stackv}. 
The surface brightness for an annulus is the average over the entire area of the annulus and is plotted
in Figure \ref{fig-stackvprofile} at the mid-point of the annulus.
The linear fit to all but the first point has a slope of $0.021\pm 0.0004$ mag pc$^{-2}$ arcsec$^{-1}$. 
This slope yields a disk scale length of 51.7\arcsec, which is 0.20 kpc at our fiducial distance of 0.8 Mpc.
This is very close to the $R_D$ of $0.19\pm0.01$ kpc for DDO 69 determined by \citet{kim13} from shallower imaging.
There is a slight rise in surface brightness level of the inner-most annulus relative to the others.
Thus, the drop off is exponential.  
\citet{d216} used star counts in DDO 216 to also trace an exponential disk to 32 \optsb\ in that galaxy.
Notably, the presence of such an exponential is found here in galaxies without spiral activity, 
suggesting that spirals are not necessary to produce or maintain the exponential profile
\citep[see also][]{zheng15}. 
Spiral structure may contribute to exponentials only in more massive galaxies
\citep{berrier}.

The extended, unbroken exponential that we find in the stacked imaging at $V$-band
suggests that we are not detecting a stellar halo because there is no change in PA or ellipticity with radius. 
On the other hand, simulations of stars scattering off of massive clumps by \citet{scatter} produced a near-exponential
disk with disk thickening. However, the galaxy in these simulations has a considerably higher stellar mass ($1.4 \times 10^9$ M\solar)
than our galaxies ($10^6$ to $6 \times10^7$ M\solar) and the individual clumps have masses that are comparable to the
total stellar mass of our galaxies ($7 \times 10^6$ M\solar). 
Thus, it is not clear how applicable these simulations would be to our \dirr\ galaxies.
Observations of a large number of massive \dirr\ galaxies by \citet{Kado} also showed most often a lack of change in $b/a$
with radius to four effective radii and broken exponential surface brightness profiles. \citet{Kado} argue that
this demonstrates in situ formation of the extended stellar component.

The simulations of \dirr\ galaxy evolution by \citet{elbadry16} found considerable dynamical evolution in small galaxies.
To find what the FIRE simulations would predict in terms of scattered stars for our galaxies, 
we looked at their Figure 9, 
which gives the average radial migration distance of stars in units of $R_{90mass}$,
the radius encompassing 90\% of the total mass of the galaxy.
For our stacked image, $R_{90mass}$ is 162\arcsec, where the $V$ surface brightness has dropped to 28.4 \optsb.
The migration distance in the simulations is a function of the total stellar mass of the model galaxy. 
Our seven galaxies have $\log M_*$ of 6 to 7.8. So, the radial migration distances are predicted
by the models to be 19\arcsec\  to 36\arcsec\ in our stacked image.
If we assume that the bulk of the stars were born within the 25 \optsb\ isophote, the stars would scatter no further, on average, than to within the
26 \optsb\ to 27 \optsb\ annulus. Thus, we would not expect the stars to scatter very far
into the far outer stellar disk and a halo should be concentrated to the center.
This could explain why we did not detect a stellar halo beyond 27 \optsb. 
The $V$ surface brightness profile does show a slight increase in surface brightness at a radius of 67\arcsec\ (26.4 \optsb), 
but the irregular contributions from star formation to the inner regions of the galaxies are likely to
overwhelm any signal from a weak stellar halo there.

\begin{figure}[t!]
\epsscale{0.8}
\plotone{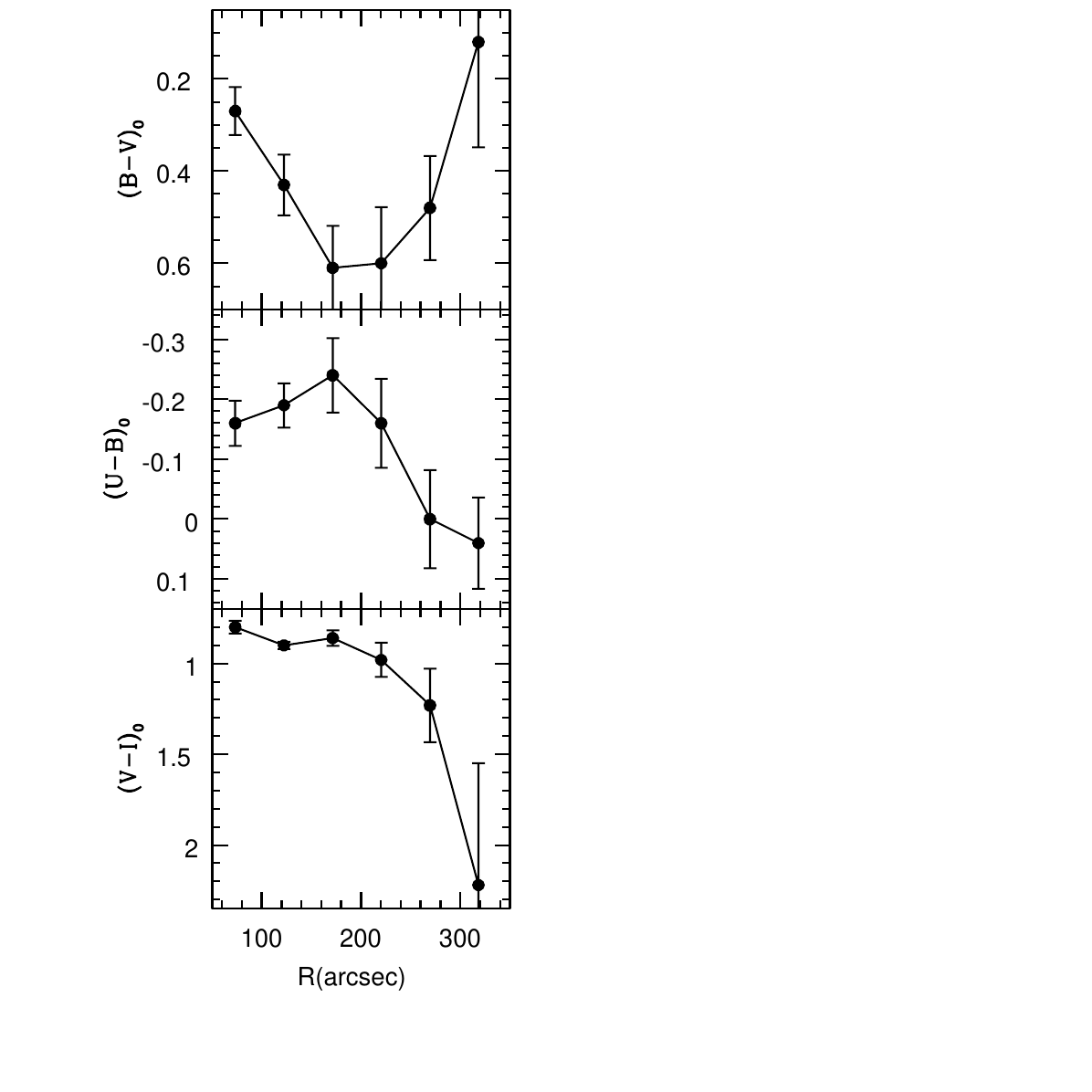}
\vskip -0.5truein
\caption{Colors of annuli from the stacked images. 
Each radius is the mid-point of the annulus.
The colors are corrected for the average foreground Milky Way reddening of the 7 stacked galaxies ($E(B-V)_f=0.02$).
The uncertainties are from a ''jackknife'' process described in the text.
\label{fig-stackcolors}}
\end{figure}

We  also stacked the $UBI$ images in order to use the colors to get approximate ages for the stellar populations
in the annuli. 
We have corrected the colors for the average foreground Milky Way reddening. The $E(B-V)_f$ of each galaxy
is given in Table \ref{tabgalaxies} and the average of the seven stacked galaxies is $E(B-V)_f = 0.02$.
The internal reddening in the central regions of \dirr\ galaxies is small, typically $E(B-V)_i = 0.05$ \citep{he06}.
This would likely be even lower in the far outer stellar disks, but actual values there are unknown. 
Therefore, we have not corrected for internal reddening.
We used the complex stellar population model of \citet{bc03} and fits to the ages as a function of color described
in Section \ref{sec-analysis} to determine ages.
The colors as a function of mid-point of each annulus is shown in Figure \ref{fig-stackcolors}\
and Table \ref{tab-ages} gives the ages determined from the colors.
We see that the ages determined from the colors agree reasonably well.
We see ages mostly of 1-6 Gyr.
In the photometry of individual stars in DDO 216, \citet{d216} found ages of 5-10 Gyr.
Similarly, \citet{zheng15} found an average outer-disk age of $\log {\rm age}\sim9.3$ for 
the far outer parts of their stacked galaxy images
of their low mass sample
($M_{\rm star}<10^{10}\:M_\odot$), although their sample is 
still $\sim 10^3\times$ more massive than our galaxies (see their Figure 8).

The outer-most annulus has an average surface brightness of 32.3 \optsb\ and a $(B-V)_0$ color of 
$0.12\pm0.04$. 
This estimates a stellar mass-to-light ratio of $0.3\pm0.2$ using the formula of \citet{kim13}
and a stellar mass surface density of 0.0013$\pm$0.0011 M\solar\ pc$^{-2}$.
These values are also similar to those in the far-outer parts of the low-mass stacked galaxy images of \citet{zheng15}.
For an average age of 3 Gyr, the star formation rate corresponding to this stellar mass density 
is $4.3\pm3.7 \times 10^{-7}$ M\solar\ pc$^{-2}$ Myr$^{-1}$.

\begin{deluxetable}{rccccccccc}
\tiny
\tablecaption{Ages of annuli. \label{tab-ages}}
\tablewidth{0pt}
\tablehead{
\colhead{$R_{mid}$} & \colhead{$(U-B)_0$} & \colhead{$(B-V)_0$} &\colhead{$(V-I)_0$} & \colhead{log age($U-B$)} 
& \colhead{log age($B-V$)} & \colhead{log age($V-I$)}
}
\startdata
 73.4 &  -0.16$\pm$0.04 &  0.27$\pm$0.05 &  0.80$\pm$0.03 &  9.20$\pm$0.02 &  9.10$\pm$0.04 &  9.34$\pm$0.37 \\
122.4 &  -0.19$\pm$0.04 &  0.43$\pm$0.07 &  0.90$\pm$0.02 &  9.14$\pm$0.02 &  9.40$\pm$0.09 &  9.61$\pm$0.46 \\
171.4 &  -0.24$\pm$0.06 &  0.61$\pm$0.09 &  0.86$\pm$0.04 &  9.03$\pm$0.03 &  9.65$\pm$0.19 &  9.51$\pm$0.42 \\
220.3 &  -0.16$\pm$0.07 &  0.60$\pm$0.12 &  0.98$\pm$0.09 &  9.20$\pm$0.02 &  9.64$\pm$0.18 &  9.79$\pm$0.54 \\
269.3 &   0.00$\pm$0.08 &  0.48$\pm$0.11 &  1.23$\pm$0.20 &  9.51$\pm$0.02 &  9.46$\pm$0.11 & 10.11$\pm$0.87 \\
318.2 &   0.04$\pm$0.08 &  0.12$\pm$0.23 &  2.22$\pm$0.67 &  9.60$\pm$0.02 &  8.56$\pm$0.02 & \nodata \\
\enddata
\end{deluxetable}

\clearpage

\section{Summary} \label{sec-summary}

We have examined the stellar structure of 10 nearby \dirr\ galaxies by fitting ellipses as a function of surface brightness.
We also compared the optical structure with the distribution of star formation as seen in FUV images.
We found the following:

\begin{enumerate}

\item The star formation activity in these dIrr galaxies is asymmetrical and patchy, which often
skews the broad-band optical isophotes. 
FUV complexes can be a large fraction of the disk scale length in size.
The stellar disks are low surface brightness and star formation
can often be asymmetrical in a way that is not usually seen in giant spiral
galaxies.

\item There are several galaxies in our sample whose morphology has been disturbed,
probably by mergers.
DDO 46 has an unusual stellar extension and
an apparent nearby tiny companion, although not all of the characteristics
of these two galaxies are consistent with an interaction between them.
DDO 187 and NGC 3738, on the other hand, both display unusual star formation morphologies
and \HI\ kinematics possibly consistent with a merger at some time in the past.

\item We looked for evidence of the presence of a stellar halo and found suggestions in
only a few galaxies either from a change in PA and/or ellipticity 
from the inner disk to the outer disk.
We explored this further by stacking the images of seven of the sample galaxies and 
reached a $V$ surface brightness of 32.3 \optsb\ in the stacked image. 
The surface photometry profile is exponential suggesting 
that we did not detect a stellar halo down to a stellar mass surface density level of 0.0013$\pm$0.0011 M\solar\ pc$^{-2}$.
Furthermore, the extended disks are likely the result of internal evolutionary processes rather than
external accretion.
$UBVI$ colors of the annuli in the stacked image
give ages of order 1-6 Gyr for the far outer stellar disk.
\end{enumerate}

\acknowledgments

We acknowledge python coding assistance from ChatGPT.
Flagstaff sits at the base of mountains sacred to tribes throughout the region. 
We honor their past, present, and future generations, who have lived here for millennia 
and will forever call this place home.

Facilities: \facility{LDT} \facility{GALEX}

\clearpage

\appendix
\section{Galaxy models constructed from isophotes} \label{models}

\begin{figure}[t]
\plotone{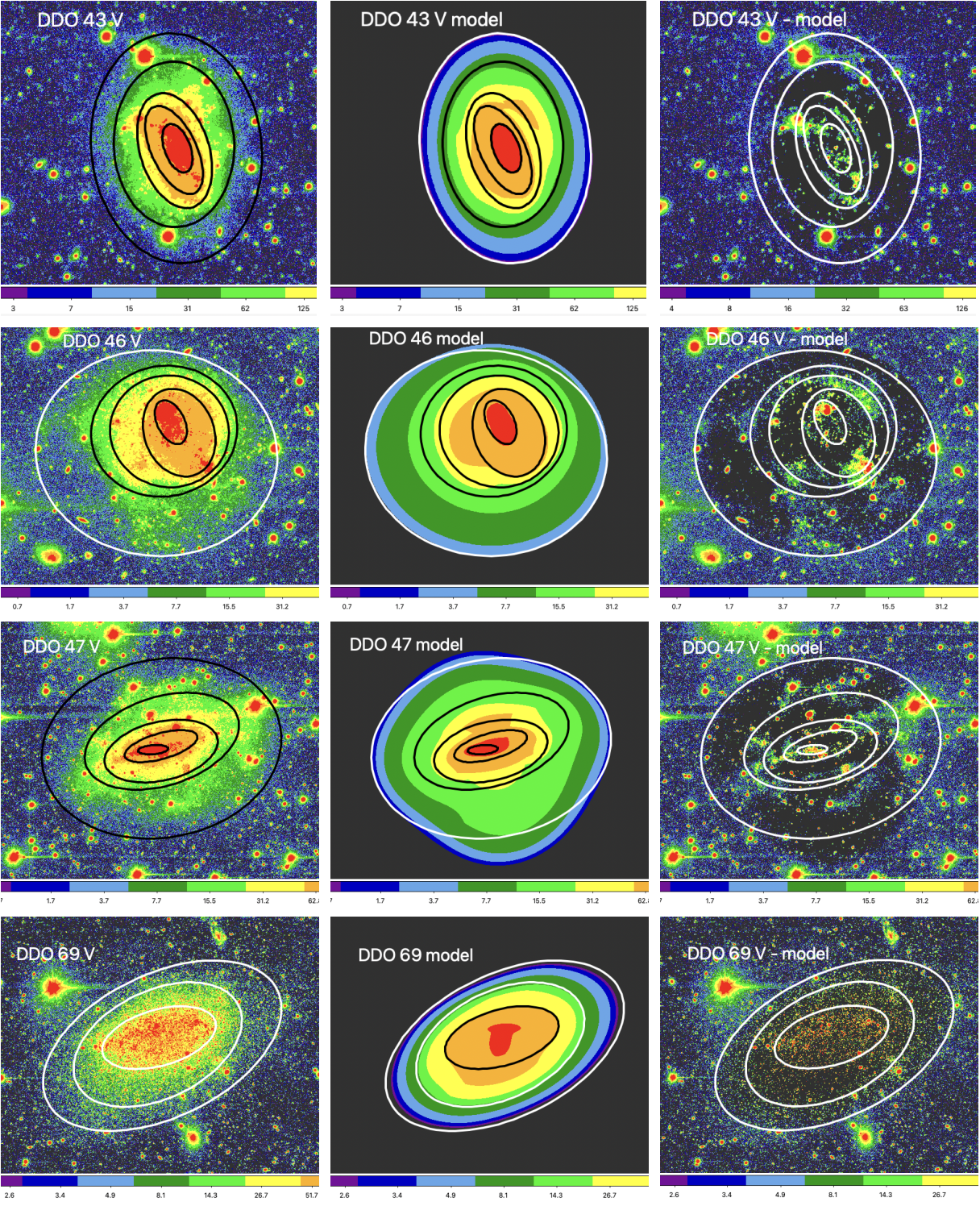}   
\caption{
All images are shown with logarithmic scale from 0 to 1000 counts.
The ellipses fit to isophotes and described in Table \ref{tabell}
are shown superposed on each image. Some ellipses are drawn in white and some in black, as necessary
for them to be visible against the background.
}
\label{fig-models1}
\end{figure}

\begin{figure}[t]
\plotone{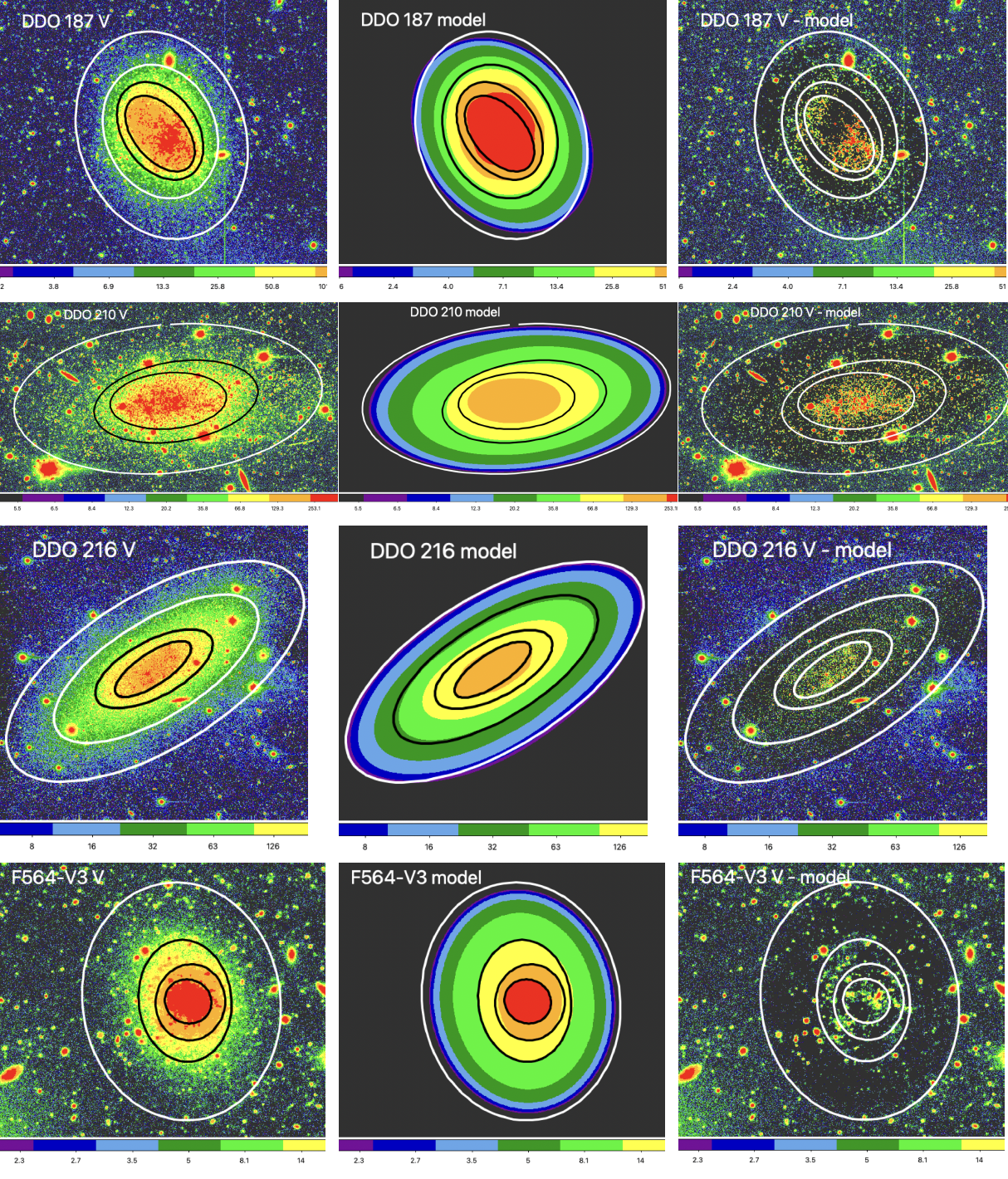}         
\caption{
As for Figure \ref{fig-models1}.
}
\label{fig-models2}
\end{figure}

\begin{figure}[t]
\plotone{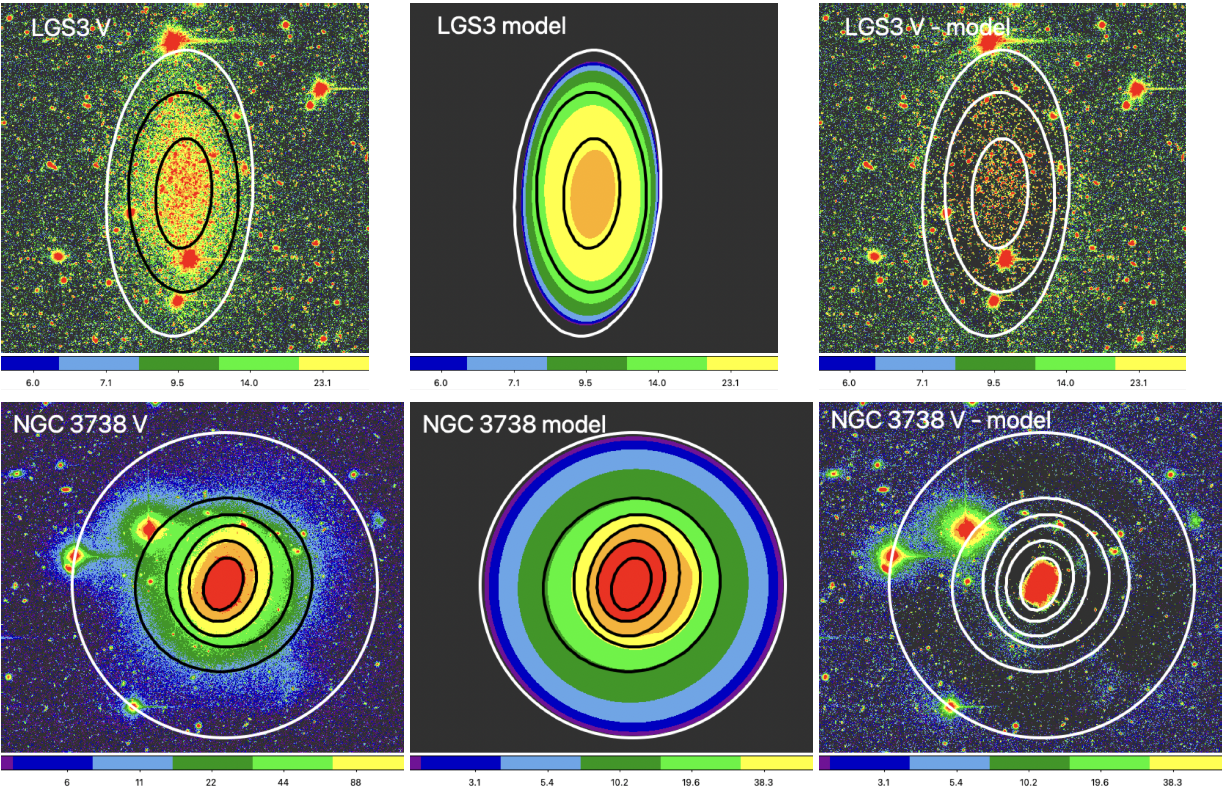}        
\caption{
As for Figure \ref{fig-models1}.
NGC 3738 is centrally concentrated, like most BCDs. As a result, the stellar brightness inside the innermost ellipse rises sharply to the center. 
This results in large residual emission interior to the first ellipse, seen in red in the difference image on the right.
}
\label{fig-models3}
\end{figure}

In Figures \ref{fig-models1}, \ref{fig-models2}, and \ref{fig-models3}
we show a comparison of the $V$ image of each galaxy (left) with a 
model image constructed from the elliptical isophotes (middle), and the difference between these two images (right). 
Since our elliptical isophotes are aimed at identifying large-scale structures, the models cannot show
small structures or deviations from elliptical structures.

\clearpage

\clearpage

\clearpage

\end{document}